\documentclass[twocolumn,prb,showpacs,citeautoscript,groupedaddress]{revtex4}
\usepackage{graphicx}
\begin{document}

\title{Evolution of optical properties of chromium spinels CdCr$_2$O$_4$, HgCr$_2$S$_4$, and ZnCr$_2$Se$_4$
 under high pressure}

\author{K. Rabia$^1$}
\author{L. Baldassarre$^1$}
\author{J. Deisenhofer$^2$}
\author{ V. Tsurkan$^{2,3}$}
\author{C. A. Kuntscher$^{1}$}\email[E-mail:~]{christine.kuntscher@physik.uni-augsburg.de}
\affiliation{$^1$ Experimentalphysik II, Universit\"at Augsburg, Universit\"atstr. 1, 86159 Augsburg, Germany}
\affiliation{$^2$ Experimentalphysik V, Universit\"at Augsburg, Universit\"atstr. 1, 86159 Augsburg, Germany}
\affiliation{$^3$ Institute of Applied Physics, Academy of Sciences of Moldova, MD2028 Chisinau, R. Moldova}

\date{\today}

\begin{abstract}
We report pressure-dependent reflection and transmission measurements on ZnCr$_2$Se$_4$, HgCr$_2$S$_4$, and CdCr$_2$O$_4$ single crystals at room temperature over a broad spectral range 200-24000~cm$^{-1}$. The pressure dependence of the phonon modes and the high-frequency electronic excitations indicates that
all three compounds undergo a pressure-induced structural phase transition with the critical pressure 15~GPa, 12~GPa, and 10~GPa for CdCr$_2$O$_4$, HgCr$_2$S$_4$, and ZnCr$_2$Se$_4$, respectively. The eigenfrequencies of the electronic transitions are very close to the expected values for chromium crystal-field transitions. In the case of the chalcogenides pressure induces a red shift of the electronic excitation which indicates a strong hybridization of the Cr d-bands with the chalcogenide bands.
\end{abstract}


\maketitle

\section{Introduction}
\begin{figure}[b]
\includegraphics[width=8.5cm]{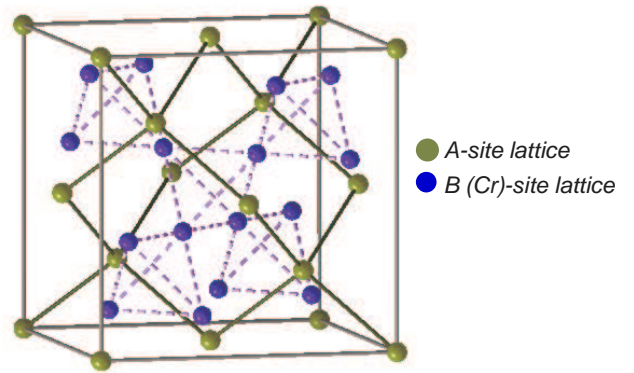}
\caption{Illustration of $A$ and $B$(Cr) sublattices in the cubic unit cell of normal spinels~\textit{AB$_{2}$X$_{4}$}. The tetrahedrally coordinated Cr-atoms form a pyrochlore lattice.}
\label{fig:ABlattice}
\end{figure}

Chromium spinel compounds with formula $A$Cr$_{2}$$X$$_{4}$, where $A$ is a divalent non-magnetic cation and $X$ a divalent anion, exhibit many interesting phenomena such as strong geometric frustration,~\cite{Lee2002, Rudolf2007,waskowska2002} relaxor multiferroicity,~\cite{Hemberger2005} or three-dimensional topological effects~\cite{Xu2011}. At room temperature these spinels are cubic (space group Fd$\bar{3}$m), where the non-magnetic A-site ions are tetrahedrally coordinated and the Cr$^{3+}$ ions (electronic configuration 3d$^{3}$ with spin S = 3/2) in octahedral environment form a pyrochlore lattice (see Fig.~\ref{fig:ABlattice}).

The inherent frustration of the pyrochlore lattice is often released by magneto-elastic
interactions such as the spin-driven Jahn-Teller effect~\cite{Yamashita2000,Tchernyshyov2002}, which lead to a reduced
structural symmetry coinciding with the onset of long-range magnetic order.
The magneto-structural phase transition is reflected in the splitting of infrared (IR) and Raman active phonons and the crystal-field (CF) excitations, which are sensitive to changes of the magnetic and the crystalline symmetry.\cite{Rudolf2007,Watanabe74,Aguilar2008,Rudolf2009,kant2009} Since magnetic interactions depend on interatomic distances~\cite{waskowska2002}, one can expect structural anomalies not only at low temperatures but also at high pressures.

In general, the application of pressure can be used to tune in a controlled fashion the electronic and structural properties of a system, while optical spectroscopy is a powerful method to investigate the electronic and vibrational properties of a material. The combination of the two techniques thus allows to follow the evolution of the phonon frequencies and the (possible) splitting of optical phonon modes with applied pressure, providing
useful information on both the magnetic ~\cite{Massidda1999,kant2009,Kant2012} and structural properties. The far-infrared studies on the spinel compounds can indeed reveal information on the structural properties (e.g., cation ordering, bond strengths and ionicities), on the free carrier contribution, and on magnetic phenomena~\cite{Lutz1983}.

Earlier work on the pressure effects in chromium spinels has concentrated on alterations of the crystal structure.\cite{waskowska2002,waskowska2004}
In this work we focus on the effect of pressure on the optical properties of ZnCr$_2$Se$_4$, HgCr$_2$S$_4$, and CdCr$_2$O$_4$ spinels in a broad frequency range (200-24000~cm$^{-1}$) at room temperature. Based on the pressure dependence of the phonon modes and the electronic excitations we propose the occurrence of a pressure-induced structural phase transition in all three compounds.

The paper is organized as follows: In Section I we discuss the experimental details of the pressure measurements and the data analysis procedure. We present the obtained optical spectra in Section II, discuss the results in Section III, and summarize our findings in Section IV.

\section{Experiment}
\label{sectionexperiment}
We have investigated the three single-crystalline Cr-spinels ZnCr$_2$Se$_4$, HgCr$_2$S$_4$, and CdCr$_2$O$_4$ by means of infrared reflection and transmission measurements under pressure in the frequency range 200-24000~cm$^{-1}$. The single crystals of ZnCr$_2$Se$_4$ and HgCr$_2$S$_4$ were grown by a chemical transport reactions \cite{Rudolf2007a,Rudolf2007}, and CdCr$_2$O$_4$ was grown by a flux method.

The pressure-dependent reflectivity and transmission measurements were performed at room temperature, using a Bruker IFS 66v/S FTIR spectrometer. Far-infrared (FIR) reflectivity measurements were carried out at beamline IR1 of ANKA. The higher frequency measurements were carried out with conventional infrared radiation sources.
A Syassen-Holzapfel diamond anvil cell (DAC) \cite{Huber1977}, equipped with type IIa diamonds suitable for infrared measurements,
was used to generate pressures up to 20 GPa.

\begin{figure}[t]
\includegraphics[width=6cm]{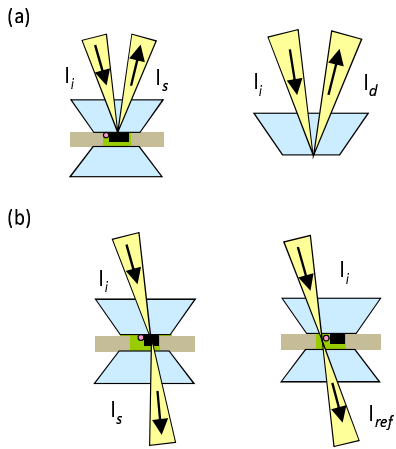}
\caption{Measurement configurations: (a) reflectance of a sample at the sample-diamond interface in a DAC and (b) transmittance of a sample in a DAC.}
\label{fig:mgeometry}
\end{figure}

A Bruker IR ScopeII infrared microscope with a 15$\times$ magnification objective coupled to the spectrometer was used to focus the infrared
beam onto the small sample in the pressure cell. The samples were mechanically polished down to a thickness of $\approx$ 50~$\mu$m and a small piece of sample (about 80~$\mu$m$\times$80~$\mu$m in size) was cut and loaded, together with CsI powder as pressure medium, in a 150 $\mu$m diameter hole drilled in a stainless steel gasket. Great care was taken when loading the sample, in order to obtain a clean and flat diamond-sample interface. With this crystal size and the corresponding diffraction limit, we were able to measure reliably in the frequency range down to 200 cm$^{-1}$. The ruby luminescence method \cite{Mao1986} was used for \emph{in-situ} pressure determination. Variations in synchrotron source intensity were taken into account by applying additional normalization procedures.

The measurement geometries are shown in Fig.~\ref{fig:mgeometry}. In case of reflectivity measurements, spectra taken at the inner diamond-air interface of the empty cell served as the reference for normalization of the sample spectra. The absolute reflectivity at the sample-diamond interface, denoted as $R_{s-d}$, was calculated according to $R_{s-d}$$(\omega)$=R$_{dia}$$\times$I$_{s}$$(\omega)$/I$_{d}$$(\omega)$, where I$_{s}$$(\omega)$ denotes the intensity spectrum reflected from the sample-diamond interface and I$_{d}$$(\omega)$ the reference
spectrum of the diamond-air interface. R$_{dia}$=0.167 was calculated from the refractive index of diamond n$_{dia}$. For obtaining the transmittance spectrum, the intensity transmitted through the sample I$_{s}$ was divided by the intensity transmitted through the pressure transmitting medium I$_{ref}$ i.e, $T$$(\omega)$=I$_{s}$$(\omega)$/I$_{ref}$$(\omega)$. The absorbance spectrum was calculated according to A$(\omega)$=-log$_{10}$T$(\omega)$.

\section{Results}
\label{sectionresults}

\begin{figure*}[t]
\includegraphics[width=12cm]{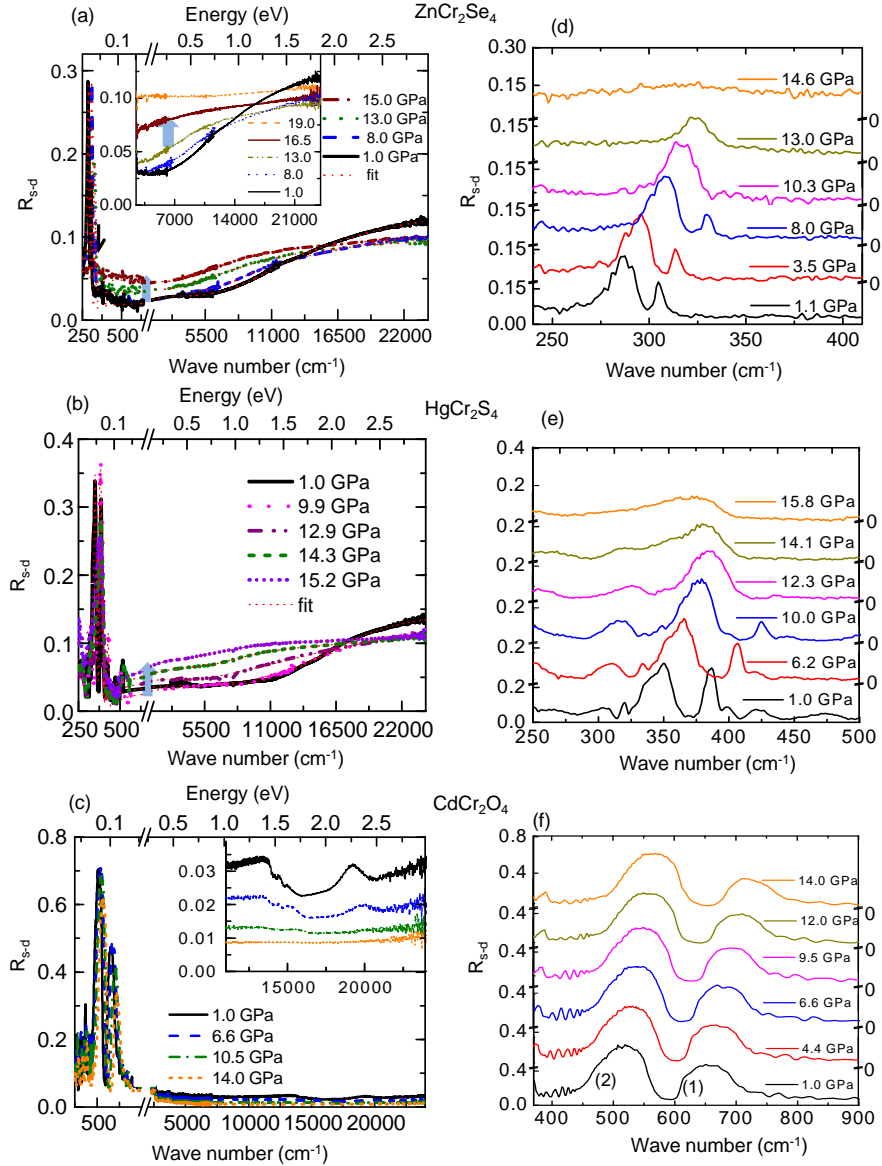}
\caption{Room-temperature reflectivity R$_{s-d}$ spectra of (a)~ZnCr$_2$Se$_4$, (b)~HgCr$_2$S$_4$, and (c)~CdCr$_2$O$_4$ in the spectral range 250-24000~cm$^{-1}$ for selected pressures. The red dashed lines are fits with the Lorentz model. In (a) the small feature at around 400~cm$^{-1}$, marked with black arrow, is an artifact and not included in the fitting. Inset: Several selected R$_{s-d}$ spectra in the pressure range 1-19.0~GPa. In (c) the inset presents the magnification of the spectra in the frequency range 11000-24000~cm$^{-1}$ to illustrate the very weak electronic excitations. The observed phonon modes of ZnCr$_2$Se$_4$, HgCr$_2$S$_4$ and CdCr$_2$O$_4$ are presented in (d), (e) and (f), respectively. The spectra are offset for clarity. The screening of the phonon modes occur for ZnCr$_2$Se$_4$, and HgCr$_2$S$_4$ with increasing pressure, while in case of CdCr$_2$O$_4$ the phonons remain intense up to the highest measured pressure.}
\label{fig:refl}
\end{figure*}

Reflectivity (R$_{s-d}$) spectra of ZnCr$_2$Se$_4$, HgCr$_2$S$_4$ and CdCr$_2$O$_4$ are shown for few selected pressures  in Fig.~\ref{fig:refl}~(a),~(b) and~(c) in the frequency range 250-24000~cm$^{-1}$. The  same curves are plotted separately in the low-frequency range [see Fig.~\ref{fig:refl}~(d)-(f)] to better highlight the effect of pressure on the phonon modes. The region around 2000~cm$^{-1}$ is cut out from the experimental spectra, due to strong diamond multi-phonon absorptions that cause artifacts in this spectral range. A linear interpolation has been performed in this range in order to perform the analysis.

For the analysis of the reflectivity data, we applied the Lorentz model, where the complex dielectric function is defined as: \cite{Gervais1983}
\begin{equation}
  \epsilon(\omega)=\epsilon_{\infty}+\sum_{j}\frac{\Delta\epsilon_{j}\omega_{TO_j}^2}
  {\omega_{TO_j}^2-\omega^2+i\omega\gamma_{j}},
  \label{eq:lorentz}
\end{equation}

In this model each mode is characterized by three parameters: the oscillator strength $\Delta\epsilon_j$,
the frequency of transverse optical modes $\omega_{TOj}$, and the damping of the modes $\gamma_j$.
The fits along with the measured reflectivity spectra are shown in Fig.~\ref{fig:refl} as red dashed lines.
The pressure dependence of the high-frequency permittivity $\epsilon_\infty$ used in our fitting was calculated according to the Clausius-Mossotti relation assuming ionic bonding:\cite{Ashcroft1976} $\frac{\epsilon_\infty(P) - 1}{\epsilon_\infty(P) + 2} = \frac{\widetilde{\alpha} N}{3 \epsilon_0 V(P)},$
%\begin{equation}
 % \frac{\epsilon_\infty(P) - 1}{\epsilon_\infty(P) + 2} = \frac{\widetilde{\alpha} N}{3 \epsilon_0 V(P)},
 % \label{eq:Clausius}
%\end{equation}
where $\widetilde{\alpha}$ is the average atomic polarizability of the unit cell, obtained from the lowest-pressure data. V(P) is the unit cell volume as a function of pressure calculated from the second-order Birch equation of state:\cite{Birch1978, Holzapfel1996}   $P(x) = \frac{3}{2}B_0 x^{-7} (1-x^2), $
where $x=[\frac{V(P)}{V(0)}]^{1/3}$. The value of the bulk modulus was assumed to be $B_0$=100~GPa for ZnCr$_2$Se$_4$ and HgCr$_2$S$_4$, and $B_0$=200~GPa for CdCr$_2$O$_4$ based on x-ray diffraction data.\cite{waskowska2002}
The obtained fit values were used for the extrapolation of the experimental reflectivity spectra to lower and higher frequencies necessary for the Kramers Kronig (KK) analysis.\cite{Plaskett1963,Pashkin06}
Since in case of HgCr$_2$S$_4$ the reflectivity data are noisy below 1000~cm$^{-1}$, the fit was used to obtain the real part of the optical conductivity.

\begin{figure}[t]
\includegraphics[width=7cm]{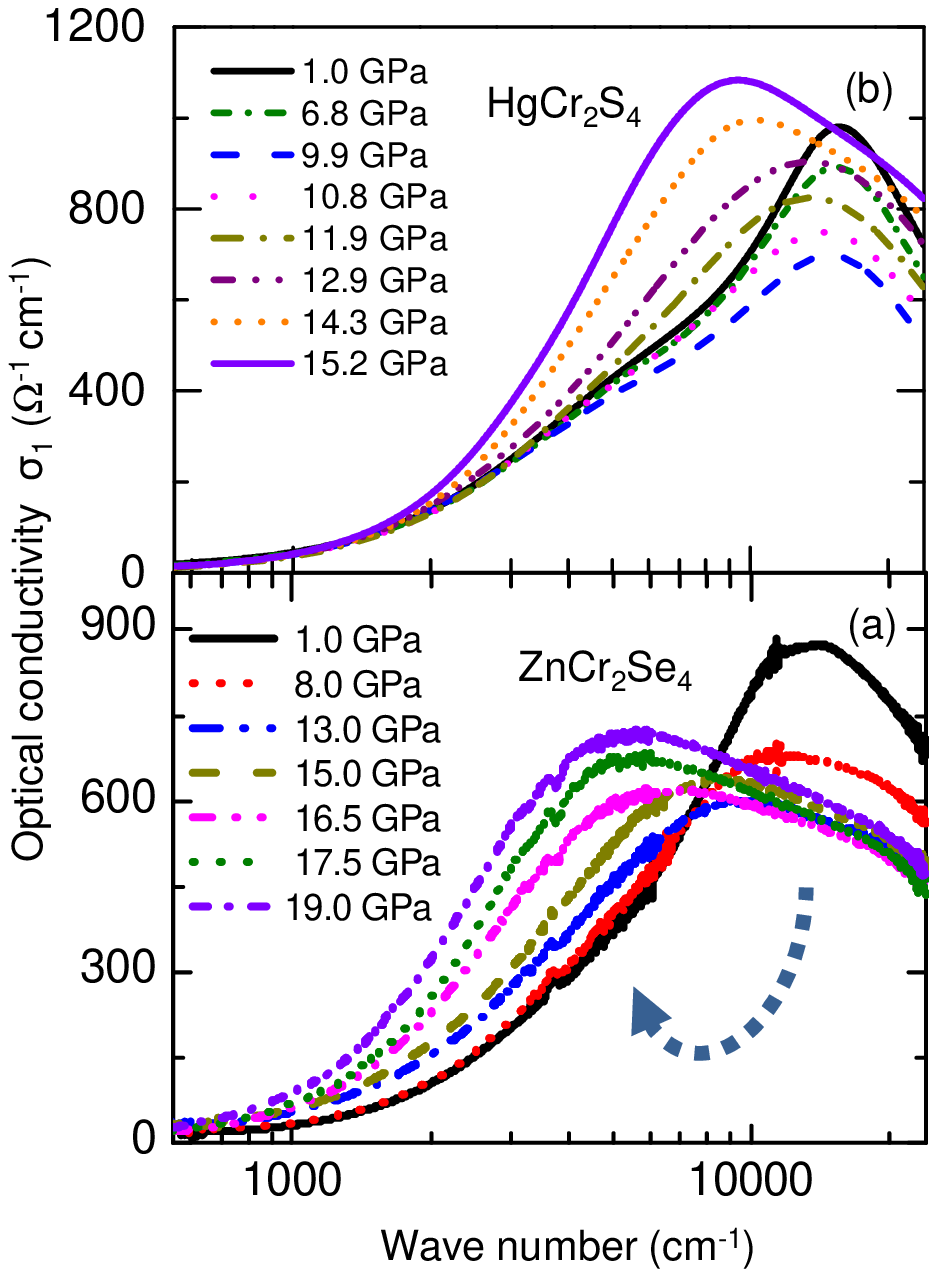}
\caption{Real part of the optical conductivity of (a) ZnCr$_2$Se$_4$ and (b) HgCr$_2$S$_4$ for selected pressures. The arrow indicates the red shift of the \textit{d-d}-transition (10~Dq).}.
\label{fig:sigma}
\end{figure}

\begin{figure}[t]
\includegraphics[width=7.5cm]{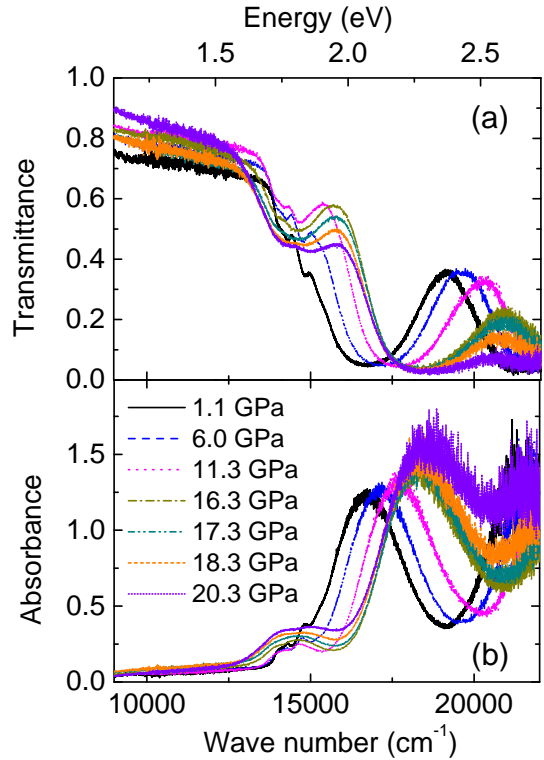}
\caption{(a) Transmittance and (b) absorbance spectra of CdCr$_{2}$O$_{4}$ for sseveral selected pressures.}
\label{fig:CdCrO_Abs}
\end{figure}

\begin{figure*}
\includegraphics[scale=1]{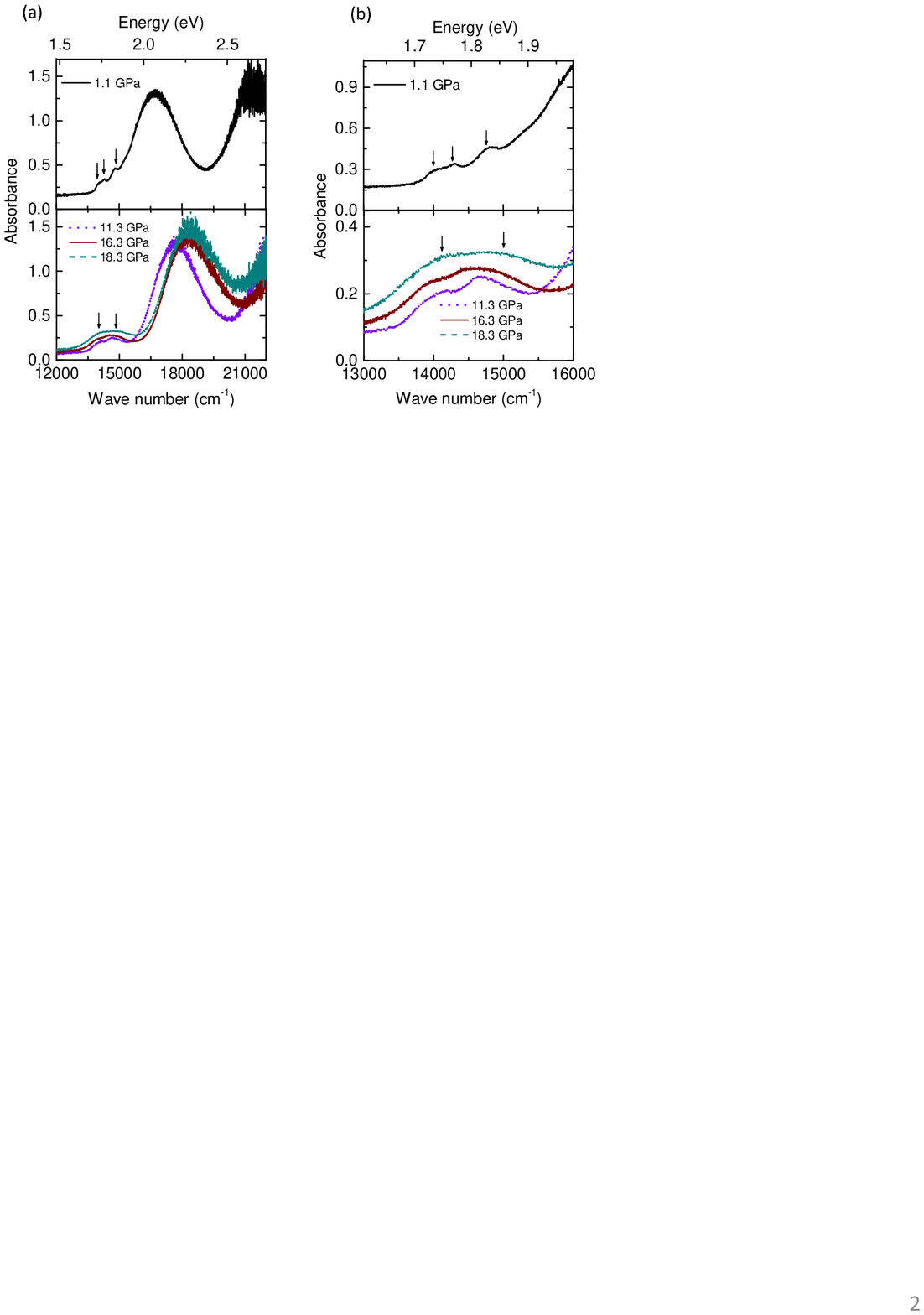}
\caption{(a) Absorbance spectrum of CdCr$_{2}$O$_{4}$ at 1.1~GPa shows four features: one strong feature at
$\approx$ 16760~cm$^{-1}$, and three weak features marked with black arrows. At higher pressures (11.3, 16.3 and 18.3~GPa) only two low-energy absorption features marked with black arrows are resolvable. (b) Low-energy part of the spectra to illustrate the pressure dependence of the spin-forbidden transitions (marked with arrows).}
\label{fig:CdCrO_Abs-1.14GPa}
\end{figure*}

\begin{figure*}
\includegraphics[scale=1]{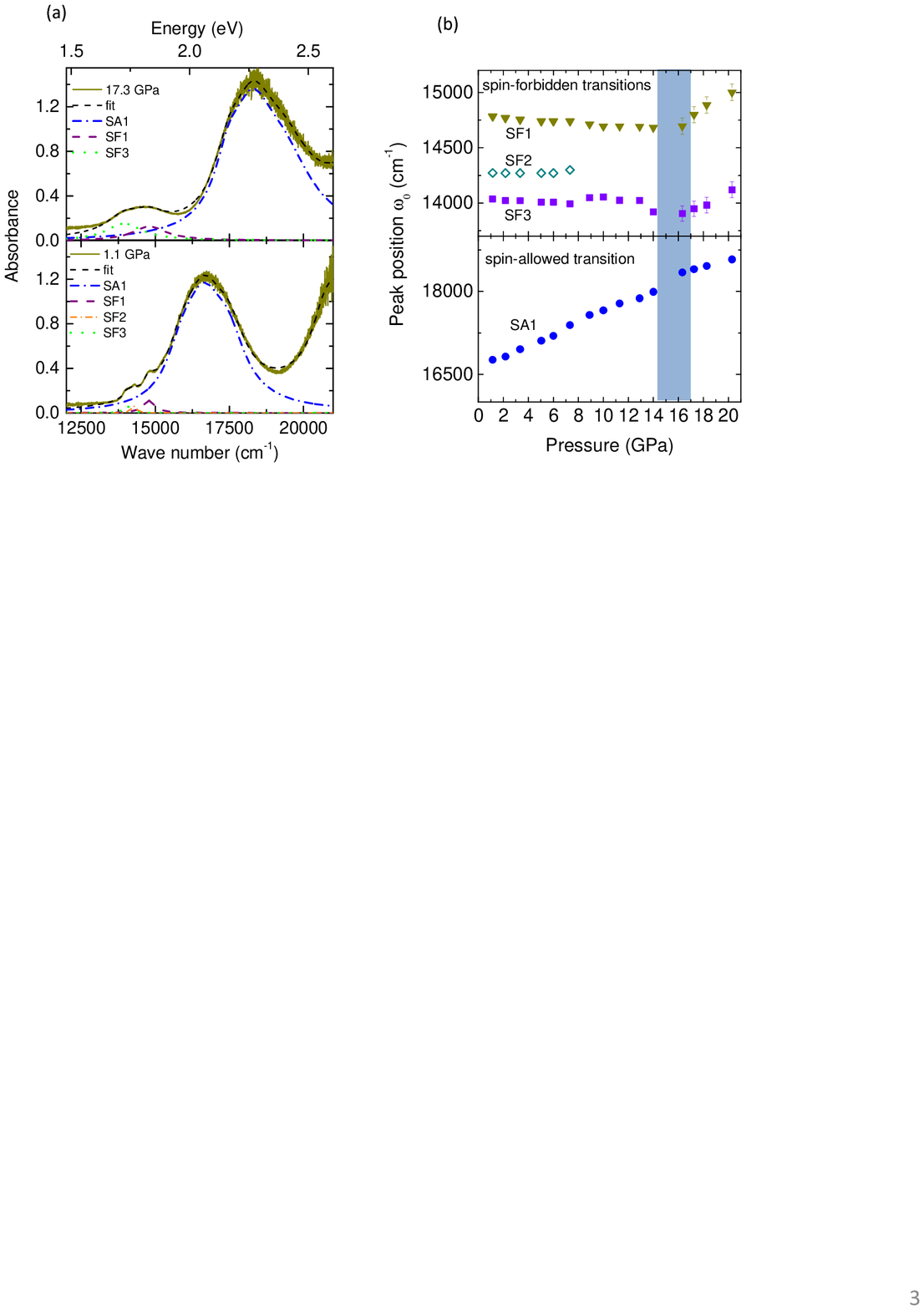}
\caption{(a) Fit of the absorption spectrum of CdCr$_{2}$O$_{4}$ at 1.1~GPa and 17.3~GPa consisting of four and three absorption features, respectively. (b) Peak position $\omega_0$ of the absorption features as a function of pressure. The shaded area marks the transition region.}
\label{fig:CdCrO_Abs-Freq}
\end{figure*}

To examine the electronic excitations in the chalcogenide compounds, the real part of the optical conductivity was considered (see Fig.~\ref{fig:sigma}). The optical conductivity spectra for ZnCr$_2$Se$_4$ and HgCr$_2$S$_4$
at the lowest pressure reveal an insulating behavior,  with two phonon modes in the low-frequency range (which are not shown in Fig.~\ref{fig:sigma}) and a strong absorption band at about 15000~cm$^{-1}$ and 15800~cm$^{-1}$, for
ZnCr$_2$Se$_4$ and HgCr$_2$S$_4$ respectively.
Upon pressure increase these bands broaden and gradually shift towards lower frequencies.

For CdCr$_{2}$O$_{4}$ the reflectivity above $\sim$5000~cm$^{-1}$ is very low and exhibits only small changes under pressure [see Fig.~\ref{fig:refl}~(c)]. Therefore, additional transmission measurements were carried out on a single crystal of CdCr$_{2}$O$_{4}$ in the frequency range 8000-25000~cm$^{-1}$ up to 20.2~GPa, especially in order to follow the effect of pressure on the high-frequency excitations.
In Fig.~\ref{fig:CdCrO_Abs} the transmittance and absorbance spectra of CdCr$_{2}$O$_{4}$ are depicted for selected pressures. These spectra show a pronounced absorption band at around 16760~cm$^{-1}$ and weak features at 14500~cm$^{-1}$. The high-frequency contributions to the absorbance spectra are shown in Fig.~\ref{fig:CdCrO_Abs-1.14GPa} for four selected pressures. The strong absorption band in CdCr$_{2}$O$_{4}$ exhibits a blue shift under pressure (see below for a quantitative analysis of the spectra).

\begin{table}[b]
\begin{center}
\footnotesize
\begin{tabular}{|c|c|}
  \hline
  Compounds & Frequency \\
  \hline
	CdCr$_{2}$O$_{4}$& 16760~cm$^{-1}$\\
 \hline
 HgCr$_{2}$S$_{4}$&15800~cm$^{-1}$\\
\hline
 ZnCr$_{2}$Se$_{4}$&15000~cm$^{-1}$\\
 \hline
\end{tabular}
\caption{The energies of the electronic transitions for the three investigated Cr-spinel compounds at $\sim$1~GPa.}
\label{tab:CFE}
\end{center}
\end{table}

In all three compounds a strong absorption bands is observed at energies (see Table~\ref{tab:CFE})
which are very close to the ones which have been attributed to spin-allowed intra-atomic d-d excitations (crystal-field excitations):\cite{Rudolf2009,Liehr1963,Jorgensen1968,Larsen1972,Golik1996,Figgis1999,Ohgushi08,Schmidt2013}
The Cr$^{3+}$ ion in an octahedral environment exhibits two spin-allowed crystal-field (CF) transitions, namely from the $^{4}$A$_{2g}$ ground state to the $^{4}$T$_{2g}$ and $^{4}$T$_{1g}$ excited states. The transition from the ground state to $^{4}$T$_{2g}$ is located at 10~Dq ($\Delta_{oct}$), where Dq is the  Coulombic parameter of the ligand field. Usually, the CF-transitions are parity-forbidden because of the inversion symmetry at the transition-metal ion site. These transitions can become allowed by virtue of lattice vibrations, which locally break the center of symmetry.\cite{Figgis1999,Rudolf2009,Schmidt2013} This interpretation is certainly fulfilled in the case of the oxide CdCr$_2$O$_4$, where these excitations can only be observed in transmission. With increasing pressure the CF-splitting ist expected to increase resulting in a blue shift of the CF excitations,\cite{Sugane1970,Kuntscher08} in agreement with our findings.

The rather large oscillator strength of the corresponding electronic excitations in the chalcogenides can only be understood if strong hybridization effects between
the chalcogenide p-states and the chromium d-states occur, and an interpretation of pure CF transition is not adequate anymore.\cite{Taniguchi1989}
The mixed character of the electronic states for the chalcogenide may also explain the observed red shift under pressure in contrast to the blue shift observed for the oxide.

In Ref.~\onlinecite{Jorgensen1968} the center of gravity of CF excitations for many chromium complexes are given, where Cr$^{3+}$ is in octahedral environment. The energies of two spin-allowed transitions from the ground state $^{4}$A$_{2g}$ to the excited levels $^{4}$T$_{2g}$ and $^{4}$T$_{1g}$ vary from $\approx$~13000-17000~cm$^{-1}$ and $\approx$17000-22000 ~cm$^{-1}$, respectively. The spin-forbidden transitions $^{4}$A$_{2g}$ to $^{2}$E$_{g}$ and $^{2}$T$_{2g}$ vary from 13000-14400~cm$^{-1}$ and 18000-19200~cm$^{-1}$, respectively.

Indeed, for CdCr$_{2}$O$_{4}$ weak features (marked by arrows) are observed at around 14500~cm$^{-1}$ in addition to the strong absorption band. They are attributed to spin-forbidden transitions, which become allowed due to spin-orbit coupling.\cite{Sugane1970} At low temperature six spin-forbidden transitions can be found, which broaden with increasing temperature and can not be completely resolved anymore at room temperature,\cite{Schmidt2013} consistent with our data.
We used Lorentz oscillators to describe the spin-allowed and spin-forbidden transitions at room temperature.
The strong absorption band at $\omega_0$$\approx$16760~cm$^{-1}$ (SA1) shows a blue shift with increasing pressure, as shown in Fig.~\ref{fig:CdCrO_Abs-Freq}~(b), but there is no significant frequency shift of the spin-forbidden excitations (SF1, SF2, SF3) up to 8~GPa. Above 8~GPa the three transitions are not distinguishable anymore, and the curves are fitted with two over-damped absorption bands. An anomalous change in their central frequencies can be seen at 14~GPa [see Fig.\ref{fig:CdCrO_Abs-Freq} (b)], showing a blue shift on further pressure increase.

\begin{figure}[t]
\includegraphics[width=7cm]{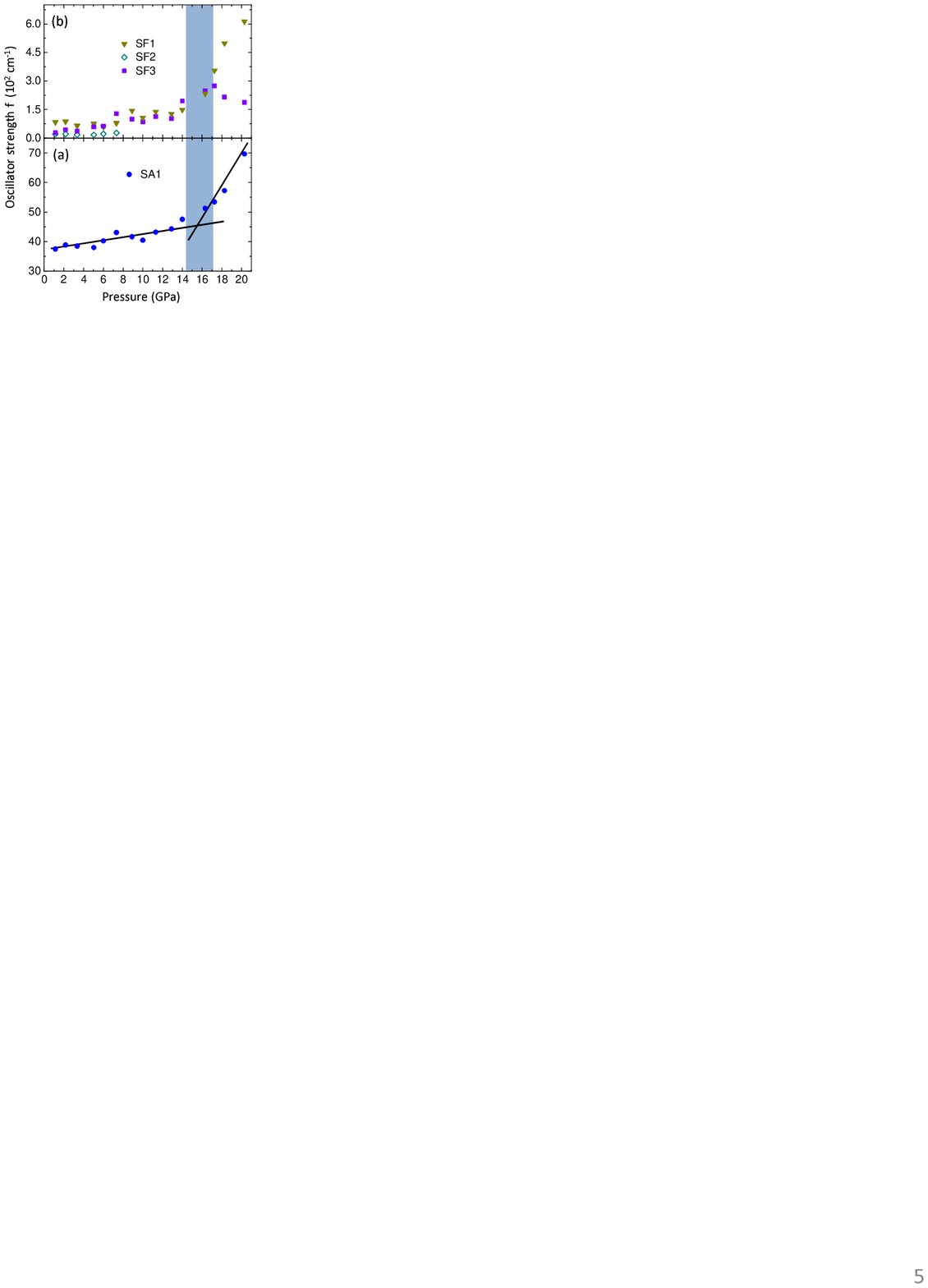}
\caption{Oscillator strength of the observed \textit{d-d}-transitions of CdCr$_2$O$_4$. (a) Oscillator strength of the spin-allowed transition (SA1) increases with increasing pressure. (b) Oscillator strength of the spin-forbidden transitions (SF1, SF2, SF3), which is very small compared to the total oscillator strength.
The shaded area marks the transition region and the black solid lines are guides to the eye.}
\label{fig:sw}
\end{figure}

\begin{figure*}
\includegraphics[scale=0.7]{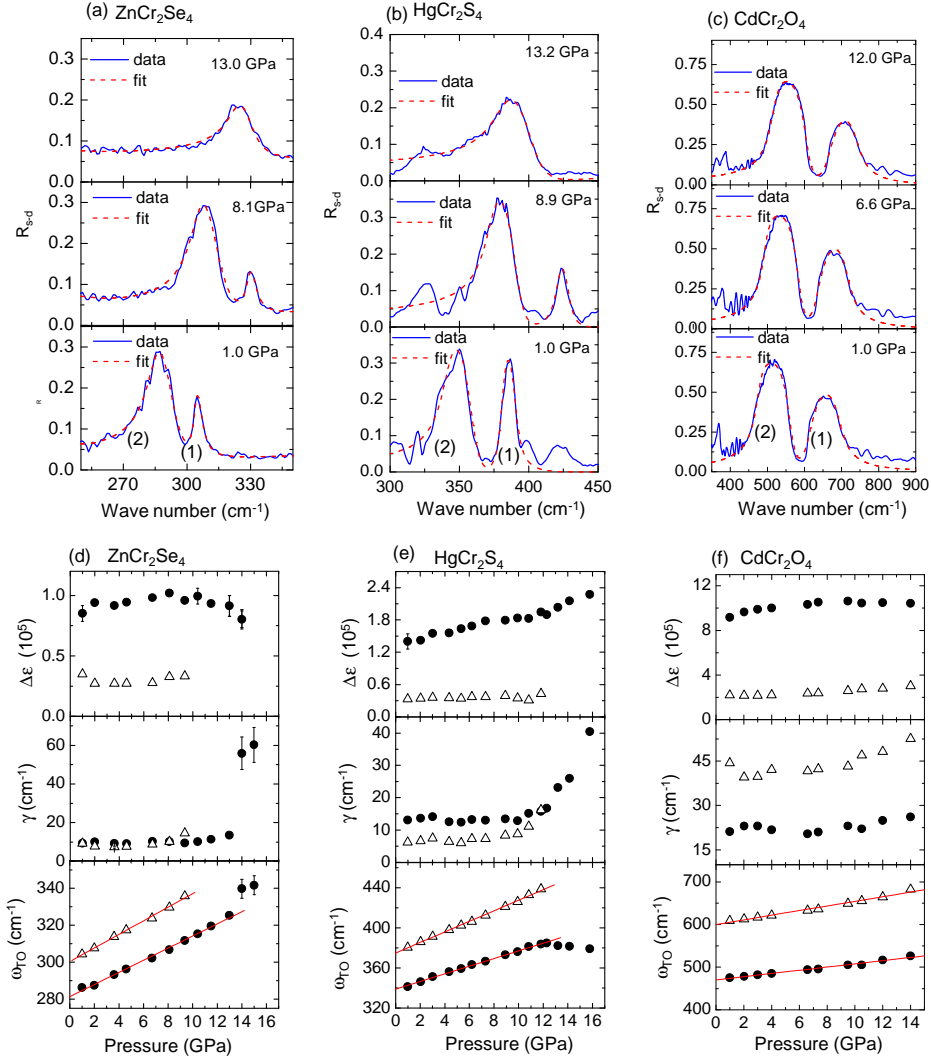}
\caption{Illustration of the fit of two observed phonon modes named as (1) and (2) with the Lorentz model according
to Eq.~\ref{eq:lorentz} for selected pressures for (a) ZnCr$_2$Se$_4$, (b) HgCr$_2$S$_4$, and (c) CdCr$_2$O$_4$. The obtained results from the fit are shown below of each relative compound (see d-f). The pressure-dependence of frequencies of optical phonons are fitted according to Eq.~\ref{eq:linear} to find the initial linear pressure coefficient.}
\label{fig:parameters}
\end{figure*}

\begin{figure*}
\begin{center}
\includegraphics[scale=0.95]{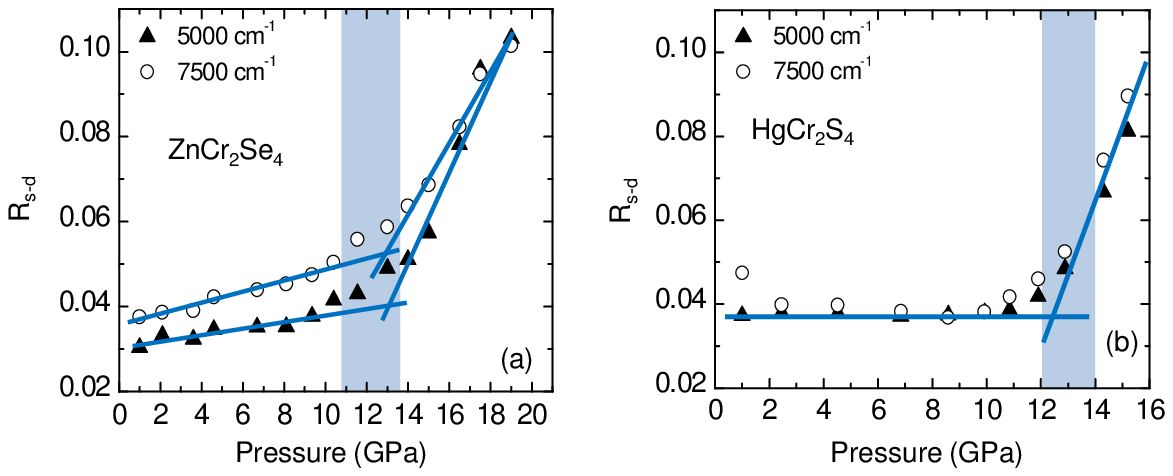}
\caption{Estimation of P$_{c}$ for (a) ZnCr$_2$Se$_4$ and (b) HgCr$_2$S$_4$ by plotting the change in reflectivity R$_{s-d}$ level with increasing pressure at 5000 cm$^{-1}$ and 7500 cm$^{-1}$ shown with solid triangles and open circles, respectively. The shaded areas mark the transition region and the full lines are guides to the eye.}
\label{fig:transpr}
\end{center}
\end{figure*}

The oscillator strength of the excitations was calculated according to
  \begin{equation}
f=\int_{\omega_1}^{\omega_{2}}A(\omega)d\omega,
  \label{eq:SW}
\end{equation}
with the absorbance $A(\omega)$, $\omega_1$=9000~cm$^{-1}$, and $\omega_2$=22000~cm$^{-1}$.
The oscillator strength of the spin-allowed transition increases with increasing pressure [see Fig.~\ref{fig:sw}(a)]. Above $P_c$$\approx$16~GPa the pressure-induced increase in oscillator strength is
drastically enhanced. The three spin-forbidden transitions contribute very little to the total oscillator strength, but overall the oscillator strength of these features increases with pressure up to $\approx$15~GPa; at this pressure anomalies occur [see Fig.~\ref{fig:sw}(b)].

In the far-infrared region we resolve two phonon modes [mode~(1) and~(2)] out of four predicted modes, since the other two low-frequency phonon modes are beyond the studied frequency range. For a more detailed discussion of the phonon modes and their evolution with pressure we refer to section IV. To extract the frequency, oscillator strength, and damping of the modes, the low-frequency reflectivity spectra were fitted with the Lorentz model. The illustration of the fit for three selected pressures for each compound and the so-obtained results for the fitting parameters are shown in Fig.~\ref{fig:parameters}.

For ZnCr$_2$Se$_4$ [see Fig.~\ref{fig:parameters}~(d)] the phonon modes harden with increasing pressure. There is no significant change in damping and oscillator strength of both phonon modes up to P$\approx$ 10~GPa. Above this pressure phonon mode~(1) is no longer visible. By further increasing the pressure, the remaining mode~(2) starts to lose its strength and broadens. At 14~GPa the phonon resonances can no longer be clearly resolved.

In the case of HgCr$_2$S$_4$, the two phonon modes harden with increasing pressure up to 12~GPa. Above 12~GPa the mode~(1) is no longer visible and the mode~(2) softens with increasing pressure. There is no significant change in oscillator strength of mode~(1), but it starts to become more damped at around 8~GPa as shown in Fig.~\ref{fig:parameters}~(e). With further increasing pressure the damping increases and the mode~(1) can no longer be resolved above 12~GPa. Mode~(2) behaves differently, since there is no significant change in the damping up to 11~GPa. By further increasing the pressure the damping increases and also the oscillator strength of mode~(2) increases. The phonon parameters show an anomaly at around 12~GPa [see Fig.~\ref{fig:parameters}~(e)].

In CdCr$_2$O$_4$ the  phonon modes harden with increasing pressure and there is no notable change in the oscillator strength of both modes~(1) and (2) up to the highest measured pressure (P $\approx$ 14~GPa) [see Fig.~\ref{fig:parameters}~(f)]. Mode~(1) loses intensity upon pressure application, but mode~(2) shows only small modifications up to the highest measured pressure.

\section{Discussion}

The optical properties of Cr-spinels have been extensively studied by various experimental methods. In literature there are numerous Raman and infrared studies reporting about the phonon frequencies for various pure and mixed spinels.
From the lattice symmetry of the normal spinels, the group theory analysis predicts four infrared-active triply degenerate T$_{1u}$  phonon modes in the FIR spectrum:\cite{Wakamura1976} \\

$\Gamma = 4T_{1u} \hspace{4.5cm} (IR-active)\\* +A_{1g}+E_{1g}+3T_{2g} \hspace{2.5cm} (Raman-active)\\* +2A_{2u}+2E_{u}+T_{1g}+2T_{2u}\hspace{2cm} (Silent)\\$

We have chosen to label  the four IR-active modes as~(1) to~(4), starting from high to low frequencies. The origin of the T$_{1u}$  modes has been the subject of controversy as two possible interpretations were proposed: In one case the two high-frequency phonon modes are related to the displacement of the Cr-\textit{X} bond in the Cr\textit{X$_{6}$} octahedra, while the two low-frequency phonon modes to the tetrahedra~\cite{Basak1994,Rudolf2007a,Rabe2005}. It has been shown experimentally~\cite{Preudhomme1971,Thirunavukkuarasu2006,Haumont2009} that the materials having an octahedral sublattice as one of the structural units exhibit phonon modes belonging to octahedral vibrations at higher-frequencies compared to the other lattice vibrations.

On the other hand,  Lutz and coworkers~\cite{Kringe2001,Lutz1999,Zwinscher1995,Zwinscher1995a,Himmrich1991,Lutz1990} claimed that almost all atoms contribute to the four IR-active phonon modes. They proposed that the contribution of \textit{A-}site atoms to the higher-frequency phonon modes (mode~(1) and (2)) is smaller compared to that to the two lower-frequency phonon modes. An assignment of the phonon modes to vibrations of tetrahedral \textit{AX$_{4}$}, octahedral\textit{ BX$_{6}$} or cubic \textit{B$_{4}$X$_{4}$} units of the structure is, within this framework, not possible~\cite{Lutz1990}. However, one can hypothesize that for higher-frequency phonon modes the contribution of octahedral vibrations is larger compared to that of tetrahedral vibrations.

It has been predicted theoretically that in normal spinels the octahedral \textit{BX$_{6}$} unit is much more ionic than the tetrahedral \textit{AX$_{4}$}, and the value of the force constant for Cr-\textit{X} is larger than that of \textit{A-X}~\cite{Zwinscher1995,Zwinscher1995a}. In infrared experiments the oscillator strength is directly related to the ionicity of a bond~\cite{Wakamura1990}. The more ionic the character of the bond is, the stronger will be the vibrational mode belonging to that bond. The mode~(2) is stronger compared to mode~(1), indicating a larger contribution of \textit{BX$_{6}$} octahedra [see  Fig.~\ref{fig:parameters}~(d)-(f)].
Furthermore, the mode intensities decrease moving from the oxide over the sulfide to the selenides, which suggests an increasing covalency of the bonds.\cite{Rudolf2007}

With increasing pressure all phonon modes shift to higher frequencies in a linear fashion [see  Fig.~\ref{fig:parameters}~(d)-(f)]. The linear pressure coefficient, \textit{C}, in the low-pressure regime was obtained by fitting the frequency of the phonon modes with the following equation:\\
\begin{equation}
\omega(P)= \omega_{o}+ C\times P,
  \label{eq:linear}
\end{equation}
where P is the applied pressure, $\omega_{o}$ is the phonon frequency at zero pressure and \textit{C} is the linear pressure coefficient. The obtained results are given in Table~\ref{tab:B}.  The value of \textit{C} represents the stiffness of bonds, and therefore can serve as a measure for the compressibility of the compound. In all investigated Cr-spinel compounds, the value of \textit{C} for mode~(2) is smaller compared to mode~(1), which attributes to the fact that the respective bonding of mode~(2) is stiffer compared to mode~(1).
Moreover the \textit{C} values for the phonon modes of CdCr$_{2}$O$_{4}$ and HgCr$_{2}$S$_{4}$ are larger than that of ZnCr$_{2}$Se$_{4}$.

\begin{table}[b]
\begin{tabular}{|c|c|c|}
  \hline
  Compounds & Phonon mode~(1) &  Phonon mode~(2) \\
  \hline
	CdCr$_{2}$O$_{4}$&5.41$\pm$0.25&3.75$\pm$0.17\\
 \hline
 HgCr$_{2}$S$_{4}$&5.25$\pm$0.08&3.80$\pm$0.06\\
\hline
 ZnCr$_{2}$Se$_{4}$&3.68$\pm$0.10&3.28$\pm$0.07\\
 \hline
\end{tabular}
\caption{Values of the linear pressure coefficient \textit{C}~(cm$^{-1}$/GPa) for two phonon modes of the three investigated Cr-spinel compounds.}
\label{tab:B}
\end{table}

Regarding the compressibility difference for the three investigated compounds, the foremost consideration is the part of the unit cell volume, which is mainly occupied by anions. The compressibility is mainly determined by the X-sublattice,\cite{waskowska2002,waskowska2004} since the number of \textit{X-X} bonds is larger compared to the other bonds (i.e., Cr-\textit{X} and \textit{A-X} bonds).
In the periodic table, the electronegativity decreases from oxygen to selenium, while the size of ions increases. The selenium (Se$^{2-}$) ion is bigger as compared to sulphur (S$^{2-}$) and oxygen (O$^{2-}$) and this corresponds to a larger bond length of chalcogenides compared to oxides. Application of external pressure will affect the longer bond length stronger than the shorter bond length. With these considerations, the chalcogenides are expected to be more compressible as compared to the oxides.  The higher compressibility corresponds to a more covalent character of the chemical bond and a lower ionicity of the anions.\cite{waskowska2002} ZnCr$_2$Se$_4$ shows semiconducting character, while CdCr$_2$O$_4$ shows insulating behavior as covalency increases from the oxide to the selenide. In this respect, the rather small values of the linear pressure coefficient \textit{C} for the modes in ZnCr$_{2}$Se$_{4}$ are surprising.

For the investigated chalcogenide compounds the phonon parameters exhibit anomalies in their pressure dependence:
For ZnCr$_2$Se$_4$ anomalies occur at P$_c$$\approx$10~GPa, which are interpreted in terms of a structural phase transition from cubic to tetragonal symmetry. This is supported by an X-ray powder diffraction study~\cite{waskowska2002} where a pressure-induced structural phase transformation from cubic (Fd$\overline{3}$m) to tetragonal (I$\overline{4}$) was reported for CdCr$_{2}$Se$_{4}$ at around 10 GPa. The phonon parameters of HgCr$_2$S$_4$ show an anomaly at around 12~GPa, which is supported by a very  recent Raman study of HgCr$_2$S$_4$ under pressure. A splitting of vibrational Raman modes was observed starting at around 12~GPa and related to a structural phase transition from a cubic to a tetragonal symmetry observed by diffraction methods.\cite{Efthimiopoulos13}
The compressibility of oxides is low compared to the chalcogenides, since the chemical bonds are very stiff, and therefore the critical pressure is expected to be higher. Indeed, a Raman study of ZnCr$_2$O$_4$ under pressure reveals a sluggish structural phase transition starting at 17.5 GPa, which is completed at around 35 GPa.\cite{Wang2002}
Consistent with this is the absence of an anomaly for the phonon modes in CdCr$_2$O$_4$ up to 14.0~GPa according to our data.

Signatures of the pressure-induced structural phase transition are also expected in the pressure evolution of the electronic excitations. Therefore, we plot in Fig.~\ref{fig:transpr} the change in the reflectivity spectra as a function of pressure at 5000 cm$^{-1}$ and 7500 cm$^{-1}$ for ZnCr$_2$Se$_4$ and HgCr$_2$S$_4$, since the reflectivity in this range relates to the electronic excitations in the materials. For ZnCr$_2$Se$_4$ the reflectivity versus pressure curve shows a continuous increase in R$_{s-d}$ with increasing pressure up to 10~GPa. By further increasing pressure there is drastic increase in R$_{s-d}$ indicating that the new structural phase is more susceptible to external pressure. In case of HgCr$_2$S$_4$ there is a drastic increase in R$_{s-d}$ above 12~GPa; below this pressure the reflectivity level remains almost constant at selected frequencies [see Fig.~\ref{fig:transpr}(b)].
CdCr$_2$O$_4$ shows an anomaly in the oscillator strength of the CF excitations at $P_c$$\approx$15~GPa
[see Figs. \ref{fig:CdCrO_Abs-Freq}(b) and \ref{fig:sw}].
A higher value of the critical pressure for CdCr$_2$O$_4$ as compared to the chalcogenides is consistent with our low-frequency results.

\section{Summary}
\label{summary}
The Cr-spinel compounds ZnCr$_{2}$Se$_{4}$, HgCr$_{2}$S$_{4}$, and CdCr$_{2}$O$_{4}$ were studied by optical spectroscopy under pressure.
With increasing pressure all observed phonon modes shift linearly to higher frequencies in the low-pressure regime.
For ZnCr$_{2}$Se$_{4}$ and HgCr$_{2}$S$_{4}$ the electronic excitations exhibit a red shift upon pressure application, whereas for CdCr$_{2}$O$_{4}$ they show a blue shift.
According to the anomalies found in the phonon behavior, reflectivity level, and the electronic excitations all three
compounds undergo a pressure-induced structural phase transition with the critical pressure $P_c$$\approx$10, 12, 15~GPa for ZnCr$_2$Se$_4$, HgCr$_2$S$_4$, and CdCr$_{2}$O$_{4}$, respectively.

\subsection*{Acknowledgements}
We acknowledge the ANKA Angstr\"omquelle Karlsruhe for the provision
of beamtime and thank B. Gasharova, Y.-L. Mathis, D. Moss,
and M. S\"upfle for assistance using the beamline ANKA-IR.
Financial support by the Bayerische Forschungsstiftung and the DFG (Emmy Noether-program, SFB 484, TRR 80)
is acknowledged.

\end{document}